\RequirePackage{ifpdf}
\ifpdf 
\documentclass[pdftex]{sigma}
\else
\documentclass{sigma}
\fi

\usepackage{bbm}
\usepackage{mathrsfs}
\usepackage{color}

\begin{document}

\allowdisplaybreaks

\renewcommand{\PaperNumber}{041}

\FirstPageHeading

\renewcommand{\thefootnote}{$\star$}

\ShortArticleName{Phase Space of Rolling Solutions of the Tippe Top}

\ArticleName{Phase Space of Rolling Solutions of the Tippe Top\footnote{This paper
is a contribution to the Vadim Kuznetsov Memorial Issue
`Integrable Systems and Related Topics'. The full collection is
available at
\href{http://www.emis.de/journals/SIGMA/kuznetsov.html}{http://www.emis.de/journals/SIGMA/kuznetsov.html}}}

\Author{S. Torkel GLAD~$^\dag$, Daniel PETERSSON~$^\dag$ and Stefan RAUCH-WOJCIECHOWSKI~$^\ddag$}

\AuthorNameForHeading{S.T.~Glad, D.~Petersson and S.~Rauch-Wojciechowski}

\Address{$^\dag$~Dept. of Electrical Engineering, Link\"opings Universitet
SE-581 83 Link\"oping, Sweden}

\EmailD{\href{mailto:torkel@isy.liu.se}{torkel@isy.liu.se}, \href{mailto:petersson@isy.liu.se}{petersson@isy.liu.se}}

\Address{$^\ddag$~Department of Mathematics, Link\"opings Universitet, SE-581 83 Link\"oping, Sweden}
\EmailD{\href{mailto:strauc@mai.liu.se}{strauc@mai.liu.se}}

\ArticleDates{Received September 15, 2006, in f\/inal form February
05, 2007; Published online March 09, 2007}

\Abstract{Equations of motion of an axially symmetric sphere rolling and sliding
on a plane are usually taken as model of the tippe top.
We study these equations in the nonsliding regime both in the vector
notation and in the Euler angle variables when they admit
three integrals of motion that are linear and quadratic in momenta.
In the Euler angle variables $(\theta,\varphi,\psi)$ these integrals give separation
equations 
that have the same structure as the equations
of the Lagrange top. It makes it possible to describe the whole
space of solutions by representing them in the space of parameters
$(D,\lambda,E)$ being constant values of the integrals of motion.}

\Keywords{nonholonomic dynamics; rigid body; rolling sphere; tippe
top; integrals of motion}

\Classification{70E18; 70E40; 70F25; 70K05}

\section{Introduction}

The tippe top (TT) has the shape of a truncated sphere with a
knob. The tippe top is well known for its counterintuitive behavior that
after being launched with suf\/f\/iciently fast spin it turns
upside down to spin on the knob until loss of energy makes it fall
down again onto the spherical bottom.

It is well known now that the sliding friction is the only force
producing a vertical component of the torque, which is needed for
reducing the vertical component of the spin and for transferring
the rotational energy into the potential energy by raising the
center of mass (CM) and inverting the tippe top.

Equations describing rolling motions of an axially symmetric sphere
are a limiting case of the TT equations and their solutions cannot
display the TT rising phenomenon since the sliding friction is absent.
They do, however indicate how solutions of the TT equations with weak
friction behave in shorter time periods. According to experiments and
numerical simulations~\cite{cohen,ebenfeld,bourabee} the rising of CM of TT
has a wobbly character which means that the inclination angle of the symmetry
axis $\theta(t)$ is rising in an oscillatory manner. The analysis
of the purely rolling solutions presented here supports an understanding that rising
of CM may be seen as a superposition of a generic nutational motion of a rolling
sphere combined together with drift of the symmetry axis caused by the action of the frictional component of the torque.

The TT is modelled by a sphere of mass $m$ and radius $R$ having
axially symmetric distribution of mass with the center of mass
shifted along the symmetry axis by $\alpha R$ ($0<\alpha<1$)
w.r.t.\ the geometrical center~$O$.

Full equations of the rolling and sliding TT admit an angular momentum type
integral of motion $\lambda = -\boldsymbol L \boldsymbol a$ called the Jellett's integral,
where $\boldsymbol L$ is the angular momentum w.r.t.~CM and $\boldsymbol a$ is a
vector pointing from CM toward the point A of contact with
the supporting plane.
The energy of the TT has, under assumption that the friction force
$\boldsymbol F_f$ acts against the direction of the sliding velocity $\boldsymbol v_A$,
negative time derivative $\dot E = \boldsymbol v_A \boldsymbol F_f(\boldsymbol v_A)<0$ and
is decreasing monotonously. These two features of the TT equations allow,
under some additional assumptions about the reaction force, for complete
description of all asymptotic motions of the TT and for
analysis of their stability~\cite{karapet2,ebenfeld,rauchglad,bourabee}.
The asymptotic solutions of the TT constitute an invariant manifold
satisfying the conditions
$\boldsymbol v_A = \boldsymbol 0$ and $\dot{\boldsymbol v}_A=\boldsymbol 0$ and it consists of vertically
spinning motions and of tumbling solutions having constant inclination $\theta$ of the symmetry axis, so that
the sphere is rolling along a circle around f\/ixed center of mass.

The usual assumption about pure rolling of the TT sphere actually
changes the model so that the reaction force is dynamically determined.
Then the TT equations simplify, the energy $E$
is conserved, and the equations admit a third integral of motion $D$ already
known to Routh~\cite{routh}.
The existence of three integrals of motion reduces the TT equations to three
f\/irst order ODE's that can be solved by separation of variables in a similar
way as the equations for the Lagrange top (LT)~\cite{ll}.

Separation of equations for a rolling axially symmetric sphere is a known fact
\cite{routh,chaplygin,gray} but detailed analysis of all possible rolling motions of
the tippe top, as labeled here by integrals
$(D,\lambda,E)$, doesn't seem to be available in literature.
A more general discussion of the Smale diagram for rolling solutions
of an ellipsoid of revolution has been presented on \cite{zob}.
A qualitative analysis of motion of a solid of revolution on
an absolutely rough plane has been also performed in \cite{mosh} by starting
from canonical equations with nonholonomicity term and by referring to properties of the monodromy matrix.
These results, when specialised to the case of sphere, lead to the same picture as presented here.

A dif\/ferent discussion (than presented here), of the main separation equation $\dot\theta^2=f(\theta)$
and of the dependence of a modif\/ied ef\/fective potential on the Jellett's integral $\lambda$
has been recently given in \cite{gray}.
Authors of \cite{gray} also explain how the set of admissible (physical) trajectories depends
on the nonsliding condition for the components of the reaction force.

In this paper, for completness of exposition, we discuss again integrals of motion
formulated in a suitable way both in the vector notation
as well as expressed through the Euler angles $(\theta,\varphi,\psi)$.
We use them to reduce the equations of motion to the separated form $\dot\theta^2 =
f(\theta,D,\lambda,E)$. The function $f(\theta,D,\lambda,E)$ is a complicated
rational function of $z = \cos\theta$ and we study solutions of this
equation by distinguishing special types of motions that are explained
by revoking similarities with the LT.

All dynamical states of the rolling tippe top (rTT) are
illustrated as a set in
the space of parameters $(D,\lambda, E)$. This set is bounded from below by the surface
of minimal value of the energy function.

The advantage of representation of dynamical states as points in
the space $(D,\lambda,E)\in \mathbb{R}^3$ is that the vector
connecting two points can be used as a measure of distance between
two states and its components can be given physical interpretation
in terms of the energy dif\/ference and the angular momentum
dif\/ference. This can be translated into the moments of force and
the work needed for transferring the TT from one state to another.

\section{Vector equations of the rolling and sliding TT}

\begin{figure}[t]
\centerline{\includegraphics{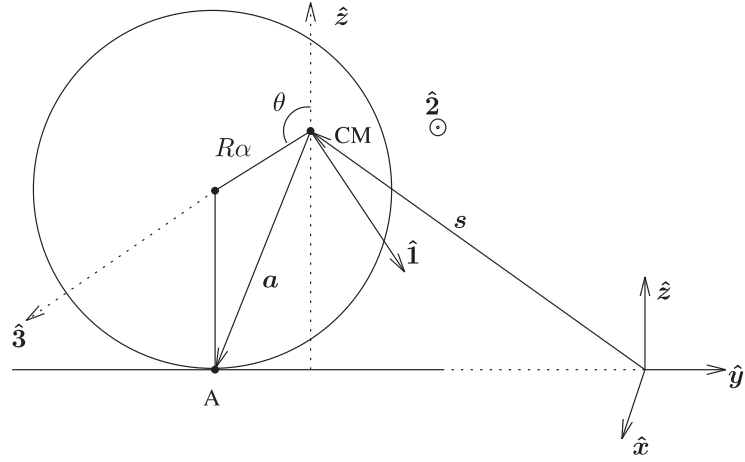}}
\caption[Model of TT]{Model of TT.}
\label{fig:TT}
\end{figure}

The tippe top (TT) is modelled by a sphere of mass $m$ and radius
$R$ having axially symmetric distribution of mass. Its center of
mass (CM) is shifted w.r.t.\ the geometric center $O$ along the
symmetry axis $\boldsymbol{\hat 3}$ by $\alpha R$, $0<\alpha<1$ as
illustrated in Fig.~\ref{fig:TT}. In describing the motion of TT
we attach at the centre of mass (CM) a moving orthonormal
reference frame ($\boldsymbol{\hat1},\boldsymbol{\hat2},\boldsymbol{\hat3}$) with the
vector $\boldsymbol{\hat3}$ directed along the symmetry axis, with the
vector $\boldsymbol{\hat1}$ (in the vertical plane of the picture)
orthogonal to $\boldsymbol{\hat3}$ and the vector $\boldsymbol{\hat2} = \boldsymbol{\hat3}
\times \boldsymbol{\hat1}$ that is always parallel to the plane of support
and is pointing out of the plane of the picture. The position of
CM w.r.t.\ the inertial orthonormal reference frame ($\boldsymbol{\hat
x},\boldsymbol{\hat y},\boldsymbol{\hat z}$) is denoted by vector $\boldsymbol s$ and the
vector connecting CM with the point~$A$, of support by the
horizontal plane, is denoted $\boldsymbol a = R\left(\alpha\boldsymbol{\hat3} -
\boldsymbol{\hat z}\right)$ as follows from Fig.~\ref{fig:TT}. The principal
moments of inertia along axis ($\boldsymbol{\hat1},\boldsymbol{\hat2},\boldsymbol{\hat3}$)
are denoted $I_1=I_2$, $I_3$ and the inertia tensor has the form
$\hat{\mathbb I} = I_1 \big\{\mathbbm 1 + \frac{I_3 - I_1}{I_1} |
\boldsymbol{\hat 3} \rangle \langle \boldsymbol{\hat 3} | \big\}$, $\hat{\mathbb
I}^{-1} = \frac{1}{I_1} \big\{\mathbbm 1 -
\frac{I_3 - I_1}{I_3} | \boldsymbol{\hat 3}\rangle\langle \boldsymbol{\hat 3} | \big\}$.

\begin{figure}[t]
\centerline{\includegraphics{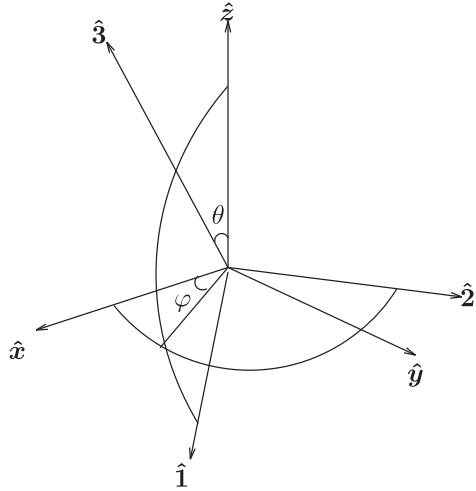}}
\caption[Axis orientation in TT]{Axis orientation in TT.}
\label{fig:coordTT}
\end{figure}

The orientation of the moving reference frame
($\boldsymbol{\hat1},\boldsymbol{\hat2},\boldsymbol{\hat3}$) w.r.t.\
the inertial reference frame
($\boldsymbol{\hat x},\boldsymbol{\hat y},\boldsymbol{\hat z}$) is described by
two angles ($\theta,\varphi$), as in Fig.~\ref{fig:coordTT}, and
the angular velocity of the moving frame can be read from Figs.~\ref{fig:TT}
and \ref{fig:coordTT}
\begin{equation*}
\boldsymbol \omega_{\rm ref} = -\dot\varphi\sin\theta\boldsymbol{\hat1} + \dot\theta\boldsymbol{\hat2} +
\dot\varphi\cos\theta\boldsymbol{\hat3},
\end{equation*}
where dots denote time derivatives of the angles ($\theta,\varphi$).
The angular velocity of TT is then
\begin{equation*}
\boldsymbol \omega = \boldsymbol \omega_{\rm ref} + \dot\psi\boldsymbol{\hat3} =
-\dot\varphi\sin\theta\boldsymbol{\hat1} + \dot\theta\boldsymbol{\hat2} +
\big(\dot\psi + \dot\varphi\cos\theta\big)\boldsymbol{\hat3}.
\end{equation*}
It contains an extra term $\dot\psi\boldsymbol{\hat3}$ that
describes rotation of TT by
the angle $\psi$ around the symmetry axis $\boldsymbol{\hat3}$;
we shall denote $\omega_3=\dot\psi+\dot\varphi\cos\theta$.
These def\/initions entail the following kinematic equations for rotation of
the reference frame ($\boldsymbol{\hat1},\boldsymbol{\hat2},\boldsymbol{\hat3}$)
\begin{gather*}
\boldsymbol{\dot{\hat1}}  =  \boldsymbol\omega_{\rm ref} \times \boldsymbol{\hat1} =
\dot\varphi\cos\theta\boldsymbol{\hat2} - \dot \theta \boldsymbol{\hat3}, \\
\boldsymbol{\dot{\hat2}}  =  \boldsymbol\omega_{\rm ref} \times \boldsymbol{\hat2} =
-\dot\varphi\cos\theta\boldsymbol{\hat1} - \dot\varphi\sin\theta\boldsymbol{\hat3}, \\
\boldsymbol{\dot{\hat3}}  =  \boldsymbol\omega_{\rm ref} \times \boldsymbol{\hat3} =
\boldsymbol\omega \times \boldsymbol{\hat3} =
\dot\theta\boldsymbol{\hat1} + \dot\varphi\sin\theta\boldsymbol{\hat2} .
\end{gather*}

The dynamics of the rolling and sliding TT is described by two ordinary
dif\/ferential equations (ODE's), one for motion of CM and
another one for rotation about CM
\begin{subequations}\label{eq:TTnewtondyn}
\begin{gather}
m\dot{\boldsymbol v}_{\rm CM} = \boldsymbol F_R + \boldsymbol F_f(\boldsymbol v_A) - mg\boldsymbol{\hat z},
 \label{eq:TTnewton}
\\
\dot{\boldsymbol L} = \boldsymbol a \times \left[ \boldsymbol F_R + \boldsymbol F_f(\boldsymbol v_A)\right],
\label{eq:TTdynamic}
\end{gather}
where $\boldsymbol v_{\rm CM} = \dot{\boldsymbol s}$, $\boldsymbol L = \hat{\mathbb
I}\boldsymbol\omega$ is the angular momentum and $\boldsymbol v_A = \boldsymbol v_{\rm CM} +
\boldsymbol\omega\times \boldsymbol a$ is the velocity of the point of contact
$A$. The gravity force $-mg\boldsymbol{\hat z}$ acts at the center of mass
and the contact force $\boldsymbol F = \boldsymbol F_R + \boldsymbol F_f(\boldsymbol v_A)$ acts at
the point $A$. It is a sum of the friction force $\boldsymbol F_f(\boldsymbol
v_A)$ parallel to the supporting plane and of the reaction force
$\boldsymbol F_R(\boldsymbol v_{\rm CM},\boldsymbol L)$ that depends on the dynamical state of
TT and does not have to be orthogonal to the supporting plane.
About the friction force we assume that it vanishes at zero
sliding velocity $\boldsymbol F_f(\boldsymbol v_A=\boldsymbol 0) = \boldsymbol 0$, but remarkably
many qualitative aspects of the motion of TT are independent of
the friction law that specif\/ies how $\boldsymbol F_f(\boldsymbol v_A)$ depends on
the contact velocity $\boldsymbol v_A$. This feature of equations
(\ref{eq:TTnewton}), (\ref{eq:TTdynamic}) is the reason why the
popular toy models of TT persistently exhibit the inverting
behavior for majority of supporting surfaces and for dif\/ferent
materials that TT is made of.

In order to close the system of vector equations
(\ref{eq:TTnewton}), (\ref{eq:TTdynamic}) we need to add the equation
\begin{equation}
\boldsymbol{\dot{\hat3}} = \frac{1}{I_1}\boldsymbol L \times \boldsymbol{\hat3}
\end{equation}
\end{subequations}
that follows from
$\boldsymbol{\dot{\hat3}} = \boldsymbol\omega \times \boldsymbol{\hat3} =
\hat{\mathbb I}^{-1} \boldsymbol L \times \boldsymbol{\hat3} = \frac{1}{I_1}
\big\{\boldsymbol L - \frac{I_3 - I_1}{I_3}\left(\boldsymbol L\boldsymbol{\hat3}\right)
\boldsymbol{\hat3}\big\}\times\boldsymbol{\hat3} = \frac{1}{I_1}\boldsymbol L \times \boldsymbol{\hat3}$
by the axial symmetry of TT.

In this paper we are studying only solutions that stay in the
supporting plane and, therefore, satisfy identically with respect
to time $t$ the algebraic condition $\boldsymbol{\hat z} \left[\boldsymbol s(t) +
\boldsymbol a(t)\right] = 0$. This condition is compatible with the
structure of equations \eqref{eq:TTnewtondyn} if all time
derivatives of $\boldsymbol{\hat z} \left[\boldsymbol s(t) + \boldsymbol a(t)\right] = 0$
are also equal to zero. The requirement of vanishing f\/irst
derivative $0 = \boldsymbol{\hat z} \big[\dot{\boldsymbol s} + \dot{\boldsymbol a} \big] =
\boldsymbol{\hat z}\left[\dot{\boldsymbol s}(t) + \boldsymbol\omega(t) \times \boldsymbol
a(t)\right] = \boldsymbol{\hat z} \boldsymbol v_A$ says that the contact velocity
$\boldsymbol v_A$ has to stay in the supporting plane all time and the
requirement of vanishing second derivative
\begin{gather*}
0  =  \boldsymbol{\hat z} \left[\ddot{\boldsymbol s}(t) +
 \frac{d}{dt}(\boldsymbol\omega(t) \times \boldsymbol a(t))\right] =
\boldsymbol{\hat z} \bigg[ \boldsymbol F_R + \boldsymbol F_f(\boldsymbol v_A) - mg\boldsymbol{\hat z} +
 m\frac{d}{dt}(\boldsymbol\omega(t) \times \boldsymbol a(t))\bigg] \\
 \phantom{0}{} =
\boldsymbol{\hat z} \left[ \boldsymbol F_R - mg\boldsymbol{\hat z} +
m\frac{d}{dt}(\boldsymbol\omega(t) \times \boldsymbol a(t))\right]
\end{gather*}
determines the vertical component of the reaction force $\boldsymbol{\hat
z} \boldsymbol F_R = mg - m\boldsymbol{\hat z}(\boldsymbol\omega(t) \times \boldsymbol a(t)\dot)$.
The planar component of $\boldsymbol F_R$ has to be def\/ined as an external
law of the reaction force or may be determined by some extra
conditions for motions satisfying \eqref{eq:TTnewtondyn}. For
instance in the model of the rising tippe top
\cite{ebenfeld,rauchglad} there has been assumed that $\boldsymbol
F_R=g_n(\boldsymbol v_{\rm CM},\boldsymbol L,\boldsymbol{\hat3})\boldsymbol{\hat z}$ is orthogonal to
the plane when $\boldsymbol F_f(\boldsymbol v_A)=-\mu g_n\boldsymbol v_A$ vanishes.

In the case of pure rolling solutions it is the rolling condition
that allows for determining the value of the total force $\boldsymbol F=
\boldsymbol F_R$ (since $\boldsymbol F_f(\boldsymbol v_A=\boldsymbol 0)=\boldsymbol 0$) so that rolling
without sliding takes place and $\boldsymbol F_R$ usually is not
orthogonal to the supporting plane.

The pure rolling solutions of TT equations have to satisfy
an additional algebraic condition $\boldsymbol v_A(t) = \boldsymbol v_{\rm CM} +
\left[\boldsymbol\omega(t) \times \boldsymbol a(t)\right] = \boldsymbol0$.
For discussing the rolling solutions of the TT equations
\eqref{eq:TTnewtondyn} we rewrite them as equations for the new
unknowns $\boldsymbol v_A$, $\boldsymbol{\hat3}$ and $\boldsymbol L$
(or $\boldsymbol\omega = \hat{\mathbb I}^{-1}\boldsymbol L$)
\begin{subequations}\label{submanifold}
\begin{gather}
 m\big[ \dot{\boldsymbol v}_A - (\boldsymbol\omega \times \boldsymbol a \dot)\big] =
\boldsymbol F_R + \boldsymbol F_f(\boldsymbol v_A) - mg\boldsymbol{\hat z}, \label{eq:submanifold1}\\
 \dot{\boldsymbol L} = \boldsymbol a \times \big[
mg\boldsymbol{\hat z} + m\dot{\boldsymbol v}_A - m(\boldsymbol\omega \times \boldsymbol a \dot)\big],
\label{eq:submanifold2} \\
 \boldsymbol{\dot{\hat3}} = \boldsymbol \omega_{\rm ref} \times \boldsymbol{\hat3} =
\boldsymbol \omega \times \boldsymbol{\hat3} = \frac{1}{I_1}\boldsymbol L \times \boldsymbol{\hat3}.
\label{eq:submanifold3}
\end{gather}
\end{subequations}
The requirement of vanishing
$\boldsymbol v_A(t) = \boldsymbol v_{CM}(t) + \left[\boldsymbol\omega(t) \times \boldsymbol a(t)\right]=\boldsymbol0$
entails that
$\dot{\boldsymbol v}_A = \dot{\boldsymbol v}_{CM} +
(\boldsymbol\omega \times \boldsymbol a\dot)=\boldsymbol0$ vanishes as well.
Then equations (\ref{eq:submanifold2}), (\ref{eq:submanifold3})
become an autonomous system of equations, with a polynomial
vector f\/ield, for $\boldsymbol L$ (or $\boldsymbol\omega$) and $\boldsymbol{\hat3}$.
They have solutions by the existence theorem for dynamical systems
and $\boldsymbol v_{CM}$
is determined from $\boldsymbol v_{\rm CM} = -\left[\boldsymbol\omega(t)\times\boldsymbol a(t)\right]$.
The condition $\dot{\boldsymbol v}_{\rm CM} = -(\boldsymbol\omega(t)\times\boldsymbol a(t)\dot)$
automatically follows and from the f\/irst equation \eqref{eq:submanifold1}
the unknown total force $\boldsymbol F = \boldsymbol F_R + \boldsymbol F_f(\boldsymbol v_A) =
m\big[-(\boldsymbol\omega\times\boldsymbol a\dot)\big] + mg\boldsymbol{\hat z}$, needed for
maintaining the pure rolling motion, can be calculated.
Thus we have shown.

\begin{proposition}
  The pure rolling constraint $\boldsymbol v_A=\boldsymbol
  v_{CM}+\boldsymbol\omega\times\boldsymbol a$ reduces equations
  \eqref{eq:TTnewtondyn} to the closed system of equations
\begin{subequations}\label{eq:proprTT}
\begin{gather}\label{eq:eq2prop1}
(\hat {\mathbb I}\boldsymbol\omega\dot) = \boldsymbol a \times \big[ mg\boldsymbol{\hat
z} - m(\boldsymbol\omega \times \boldsymbol a \dot)\big],
\\
\label{eq:eq3prop1}
\boldsymbol{\dot{\hat3}} = \boldsymbol\omega \times \boldsymbol{\hat3}
\end{gather}
\end{subequations}
for the unknowns $\boldsymbol\omega$ and $\boldsymbol{\hat3}$ where
$\hat{\mathbb I} = I_1 \big\{\mathbbm 1 + \frac{I_3 - I_1}{I_1} | \boldsymbol{\hat 3}
  \rangle \langle \boldsymbol{\hat 3} | \big\}$.
It is consistent
with equations~\eqref{eq:TTnewtondyn} when we assume that the
total force is dynamically determined as $\boldsymbol F=\boldsymbol F_R+\boldsymbol F_f=
-m\big( \boldsymbol\omega\times\boldsymbol a\dot{\big)} + mg \boldsymbol{\hat z}$.
\end{proposition}

Thus the pure rolling constraint changes the model of the TT by saying that
the force $\boldsymbol F$ applied to the body at point $A$ is dynamically
determined. This means that general rolling solutions presented
here usually do not satisfy the TT model with $\boldsymbol F_R=g_n\boldsymbol{\hat
z}$, $\boldsymbol F_f=-\mu g_z\boldsymbol v_A$ except the vertical spinning
solutions and the tumbling solutions with CM f\/ixed in space
\cite{ebenfeld,rauchglad}.

\section{Coordinate form of the rolling TT (rTT) equations.\\
Integrals of motion}

The autonomous system of rTT equations
(\ref{eq:eq2prop1}), (\ref{eq:eq3prop1})
can be expressed in the moving reference frame
$\boldsymbol{\hat1}$, $\boldsymbol{\hat2}$, $\boldsymbol{\hat3}$. Recall that $\boldsymbol{\hat z} =
-\sin\theta\boldsymbol{\hat1} + \cos\theta\boldsymbol{\hat3}$, $\boldsymbol a =
R(\alpha\boldsymbol{\hat3} - \boldsymbol{\hat z}) = R\big( \sin\theta\boldsymbol{\hat1} +
\left( \alpha - \cos\theta \right)\boldsymbol{\hat3}\big)$, $\boldsymbol \omega = \boldsymbol\omega_{\rm ref} +
\dot\psi\boldsymbol{\hat3} = -\dot\varphi\sin\theta\boldsymbol{\hat1} +
\dot\theta\boldsymbol{\hat2} + (\dot\psi +
\dot\varphi\cos\theta)\boldsymbol{\hat3} =
-\dot\varphi\sin\theta\boldsymbol{\hat1} + \dot\theta\boldsymbol{\hat2} +
\omega_3\boldsymbol{\hat3}$. By substituting these expressions into
\eqref{eq:eq2prop1} we get at $\boldsymbol{\hat1}$, $\boldsymbol{\hat2}$, $\boldsymbol{\hat3}$
the following system of equations for the Euler angles
$(\theta,\varphi,\psi)$
\begin{subequations}
\begin{gather}
I_3\omega_3\dot\theta -
2I_1\dot\varphi\dot\theta\cos\theta - I_1\ddot\varphi\sin\theta +
mR^2\left( \alpha - \cos\theta \right)\big(  - \ddot\varphi\sin\theta\left( \alpha - \cos\theta \right)  \nonumber \\
\qquad {}-2\dot\varphi\dot\theta\cos\theta\left( \alpha - \cos\theta \right) -
\dot\omega_3\sin\theta - \omega_3\dot\theta\cos\theta \big) = 0, \label{eq:firstcomp}\\
I_1\ddot\theta - I_1\dot\varphi^2\sin\theta\cos\theta +
I_3\omega_3\dot\varphi\sin\theta + mR^2\sin\theta\big(
\dot\theta^2\alpha +
\omega_3\dot\varphi\sin^2\theta +\ddot\theta\sin\theta \nonumber \\
\qquad{}
+\dot\varphi^2\sin^2\theta\left( \alpha - \cos\theta \right)\big) +
mR^2\left( \alpha - \cos\theta \right)\big( \ddot\theta\left( \alpha - \cos\theta \right) -
\dot\varphi\omega_3\sin\theta\cos\theta  \nonumber \\
\qquad {}- \dot\varphi^2\sin\theta\cos\theta\left( \alpha - \cos\theta \right) \big) = -\alpha
mgR\sin\theta, \label{eq:secondcomp} \\
I_3\dot\omega_3 + mR^2\sin\theta \big(
2\dot\varphi\dot\theta\cos\theta\left( \alpha - \cos\theta \right) +
\ddot\varphi\sin\theta\left( \alpha - \cos\theta \right) + \dot\omega_3\sin\theta
+ \omega_3\dot\theta\cos\theta\big) = 0. \!\!\!\label{eq:thirdcomp}
\end{gather}
\end{subequations}
After resolving w.r.t. ($\ddot\theta,\ddot\varphi,\dot\omega_3$)
we obtain
\begin{subequations}\label{eq:theequations}
\begin{gather}
\ddot\theta  =
\frac{\sin\theta}{I_1+mR^2\big(\left( \alpha - \cos\theta \right)^2+\sin^2\theta\big)}
\bigg[ \dot\varphi^2\big(-mR^2\left( \alpha - \cos\theta \right)\left( 1-\alpha\cos\theta\right)+ I_1\cos\theta\big)
  \nonumber \\
\phantom{\ddot\theta  =}{}  +
\omega_3\dot\varphi\left( mR^2\left(\alpha\cos\theta - 1\right) -
I_3\right) - mR^2\dot\theta^2\alpha - mR\alpha g\bigg],
\label{eq:ttmacddtheta}\\
\ddot \varphi  =
\frac{\omega_3\dot\theta}{\sin\theta}\left[
\frac{I_3^2 + mR^2I_3\left( 1-\alpha\cos\theta\right)}
{I_1I_3 + mR^2I_1\sin^2\theta + mR^2I_3\left( \alpha - \cos\theta \right)^2}\right] -
\frac{2\dot\varphi\dot\theta\cos\theta}{\sin\theta},\label{eq:ttmacddphi} \\
\dot \omega_3  =
-\omega_3\dot\theta\sin\theta\left[\frac{mR^2I_3\left( \alpha - \cos\theta \right)
+ mR^2I_1\cos\theta}{mR^2I_3\left( \alpha - \cos\theta \right)^2 + I_1I_3 +
mR^2I_1\sin^2\theta}\right].\label{eq:ttmacdomega}
\end{gather}
\end{subequations}
Since $\varphi$ and $\psi$ are cyclic coordinates,
that do not appear in the right hand side of equations \eqref{eq:theequations},
it is ef\/fectively a fourth order dynamical system for the variables
$\theta$, $\dot\theta$, $\dot\varphi$ and $\omega_3 = \dot\psi +
\dot\varphi\cos\theta$. It admits three functionally independent
integrals of motion~\cite{routh}.

Equation \eqref{eq:ttmacdomega} can be integrated directly to the Routh
integral
\begin{equation*}
D = I_3\omega_3\big[\gamma + \beta\left( \alpha - \cos\theta \right)^2 + \beta\gamma\sin^2\theta\big]^
\frac{1}{2} =: I_3\omega_3 \sqrt{d(\theta)},
\end{equation*}
where $\beta = \frac{mR^2}{I_3}$,
$\gamma = \frac{I_1}{I_3}$ and $d(\theta) = \gamma + \beta\big(\alpha - \cos\theta\big)^2 + \beta\gamma\sin^2\theta$.

The Jellett's integral
\begin{equation*}
\lambda = I_1\dot\varphi\sin^2\theta-\left( \alpha - \cos\theta \right) I_3\omega_3
\end{equation*}
follows from equations \eqref{eq:ttmacddphi} and
\eqref{eq:ttmacdomega} and the energy integral
\begin{gather*}
E  = \frac{1}{2}mR^2\big[ \dot\theta^2\left( \alpha - \cos\theta \right)^2 +
\sin^2\theta\dot\varphi^2\left( \alpha - \cos\theta \right)^2 +
2\sin^2\theta\dot\varphi\omega_3\left( \alpha - \cos\theta \right) \nonumber \\
\phantom{E=}{} + \sin^2\theta\omega_3^2 +
\dot\theta^2\sin^2\theta\big] +
\frac{1}{2}\big[ I_1\dot\varphi^2\sin^2\theta + I_1\dot\theta^2 +
I_3\omega_3^2\big] + mgR\left( 1 - \alpha\cos\theta\right)
\end{gather*}
is a consequence of all three equations \eqref{eq:theequations}
as can be checked by direct dif\/ferentiation w.r.t.\ time.

\begin{proposition}
The rTT equations of motion \eqref{eq:proprTT} admit three time
independent integrals of motion
\begin{subequations}
\begin{gather}
D  =  (\boldsymbol\omega \boldsymbol{\hat3})
\left[I_1I_3 + mR^2I_3\left(\alpha - (\boldsymbol{\hat z} \boldsymbol{\hat3})\right)^2 +
mR^2I_1\left( 1 - \left( \boldsymbol{\hat z} \boldsymbol{\hat3}\right)^2\right)\right]^\frac{1}{2}
 \nonumber \\
 \phantom{D}{}=  \omega_3\left[I_1I_3 + mR^2I_3\left( \alpha - \cos\theta \right)^2 + mR^2I_1\sin^2\theta\right]^
\frac{1}{2}, \\
\lambda  =  -\boldsymbol L \boldsymbol a = I_1\dot\varphi\sin^2\theta-\left( \alpha - \cos\theta \right) I_3\omega_3, \\
E  =  \frac{1}{2}m\boldsymbol v_{CM}^2 + \frac{1}{2}\boldsymbol\omega\boldsymbol L
+ mg\boldsymbol s\boldsymbol{\hat z} = \frac{1}{2}mR^2\big[ \dot\theta^2\left( \alpha - \cos\theta \right)^2 +
\sin^2\theta\omega_3^2  \nonumber \\
\phantom{E  =}{} + \sin^2\theta\dot\varphi^2\left( \alpha - \cos\theta \right)^2 +
2\sin^2\theta\dot\varphi\omega_3\left( \alpha - \cos\theta \right) +
\dot\theta^2\sin^2\theta\big]  \nonumber \\
\phantom{E  =}{} +\frac{1}{2}\big[ I_1\dot\varphi^2\sin^2\theta + I_1\dot\theta^2 +
I_3\omega_3^2\big] + mgR\left( 1 - \alpha\cos\theta\right).
\end{gather}
\end{subequations}
\end{proposition}

\begin{proof}
As we have seen, the coordinate form of the integrals of motion
follows easily from the coordinate equations \eqref{eq:theequations}.
It is instructive also to see how the vector form of the integrals of
motion is related to the vector form of equations \eqref{eq:proprTT}.

We see that the Jellett's integral $\lambda=-\boldsymbol L\boldsymbol a$ is a
scalar product of the angular momentum $\boldsymbol L$ and a vector vector $\boldsymbol a$.
When deriving
\begin{gather*}
-\dot\lambda  =  \big(\boldsymbol L \boldsymbol a\dot{\big)} =
\dot{\boldsymbol L}   \boldsymbol a +
\boldsymbol L   \dot{\boldsymbol a} = \left(\boldsymbol a \times \boldsymbol F\right) 
\boldsymbol a + \boldsymbol L
\big(\alpha\boldsymbol{\dot{\hat 3}} - R\boldsymbol{\dot{\hat z}}\big)\\
\phantom{-\dot\lambda}{} =
\alpha\boldsymbol L  \boldsymbol{\dot{\hat 3}}
 =  \alpha\frac{1}{I_1}\boldsymbol L \left(\boldsymbol L\times\boldsymbol{\hat3}\right) = 0
\end{gather*}
we see that each term disappears on its own. The f\/irst term
disappears because the vector $\boldsymbol a$ is the same as in the
dynamical equation \eqref{eq:eq2prop1}. The second term disappears
since $\boldsymbol{\dot{\hat3}} = \boldsymbol\omega \times \boldsymbol{\hat3} =
\frac{1}{I_1}\left(\boldsymbol L \times \boldsymbol{\hat3}\right)$.

In calculating time derivative of the energy integral we use the equality
$\dot{\boldsymbol\omega}\boldsymbol L = \boldsymbol\omega\dot{\boldsymbol L}$ that follows by
taking the derivative of
$\boldsymbol\omega = \hat{\mathbb I}^{-1}\boldsymbol L =$
$\frac{1}{I_1} \big\{\mathbbm 1 + \frac{I_3 - I_1}{I_3}
| \boldsymbol{\hat3}\rangle
\langle \boldsymbol{\hat3} | \big\}\boldsymbol L$.
Then for the general equations of TT with sliding
\eqref{eq:TTnewtondyn} we get
\begin{gather*}
\dot E  =  m\boldsymbol v_{CM}\dot{\boldsymbol v}_{CM} +
\boldsymbol\omega \dot{\boldsymbol L} + mg\dot{\boldsymbol s}  \boldsymbol{\hat z} \\
\phantom{\dot E}{}=
\boldsymbol F  \boldsymbol v_A - \left(\boldsymbol\omega\times\boldsymbol a\right) \boldsymbol F -
 \boldsymbol F  \left(\boldsymbol\omega\times\boldsymbol a\right) +
mg\left( \dot{\boldsymbol s} + \boldsymbol\omega\times\boldsymbol a\right) \boldsymbol{\hat z} -
mg\boldsymbol v_A \boldsymbol{\hat z} =
\boldsymbol F   \boldsymbol v_A.
\end{gather*}
For rTT $\boldsymbol v_A = \boldsymbol 0$ and the energy is
conserved since the force $\boldsymbol F$ doesn't perform any work.

The Routh integral follows remarkably simple from the coordinate equation
\eqref{eq:ttmacdomega} but is considerably more dif\/f\/icult to see in the
vector notation. The time derivative
\begin{equation*}
\dot D = I_3\left[ \frac{
2\left(\dot{\boldsymbol\omega} \boldsymbol{\hat3}\right) d\left(\boldsymbol{\hat z}\boldsymbol{\hat3}\right) + \left(\boldsymbol\omega \boldsymbol{\hat3}\right) d'\left(\boldsymbol{\hat z}\boldsymbol{\hat3}\right)}
{2\sqrt{d\left(\boldsymbol{\hat z}\boldsymbol{\hat3}\right)}}\right]
\end{equation*}
contains the term $\left(\dot{\boldsymbol\omega} \boldsymbol{\hat3}\right)$ that has to be eliminated with the
use of equations \eqref{eq:proprTT}. To do this we express
$\dot{\boldsymbol L}$ through $\dot{\boldsymbol\omega}$ by dif\/ferentiating $\boldsymbol L
= \hat{\mathbb I}\boldsymbol\omega$ and calculate the $\boldsymbol{\hat1}$ and
$\boldsymbol{\hat3}$ components that in the vector notation correspond to
\eqref{eq:firstcomp} and \eqref{eq:thirdcomp}. We obtain two
linear algebraic equations for the unknowns $\left(\dot{\boldsymbol\omega} \boldsymbol{\hat3}\right)$ and
$\left(\dot{\boldsymbol\omega} \boldsymbol{\hat z}\right)$. One by multiplying with $\boldsymbol{\hat3}$
\begin{gather}
  \left(\dot{\boldsymbol\omega} \boldsymbol{\hat3}\right)  =
  \frac{1}{I_3}\big(\hat3 \dot{\boldsymbol L}\big) =
  \beta\big[\left(\dot{\boldsymbol\omega} \boldsymbol{\hat3}\right)
  \left(\left(\boldsymbol{\hat z}\boldsymbol{\hat3}\right)\alpha-1\right)
 - \left(\dot{\boldsymbol\omega} \boldsymbol{\hat z}\right)
 \left(\alpha-\left(\boldsymbol{\hat z}\boldsymbol{\hat3}\right)\right) +
  \alpha\left(\boldsymbol\omega \boldsymbol{\hat3}\right)
  \left(\left(\boldsymbol{\hat{3}}\times\boldsymbol{\hat{z}}\right)
 \boldsymbol\omega\right)\big]\label{eq:deriverouth}
\end{gather}
and a second equation by multiplying with $\boldsymbol{\hat1}$ and substituting
$\boldsymbol{\hat1} = \frac{\hat3\left(\boldsymbol{\hat z}\boldsymbol{\hat3}\right)
- \hat z}{\sqrt{1-\left(\boldsymbol{\hat z}\boldsymbol{\hat3}\right)^2}}$. This equation
is more complicated but also involves $\left(\dot{\boldsymbol\omega} \boldsymbol{\hat3}\right)$
and $\left(\dot{\boldsymbol\omega} \boldsymbol{\hat z}\right)$ like equation \eqref{eq:deriverouth}.
From this linear system we determine $\left(\dot{\boldsymbol\omega} \boldsymbol{\hat3}\right)$ and substitute into
the $\dot D$ expression.
One then obtains
\begin{gather*}
\dot D  =  I_3\left[ \frac{
2\left(\dot{\boldsymbol\omega} \boldsymbol{\hat3}\right) d\left(\boldsymbol{\hat z}\boldsymbol{\hat3}\right) + \left(\boldsymbol\omega \boldsymbol{\hat3}\right) d'\left(\boldsymbol{\hat z}\boldsymbol{\hat3}\right)}
{2\sqrt{d\left(\boldsymbol{\hat z}\boldsymbol{\hat3}\right)}}\right] \\
\phantom{\dot D}{}= \frac{2I_3\left(\boldsymbol\omega \boldsymbol{\hat3}\right)\left(\left(\boldsymbol{\hat{3}}\times\boldsymbol{\hat{z}}\right)
 \boldsymbol\omega\right)\beta}
{2\sqrt{d\left(\boldsymbol{\hat z}\boldsymbol{\hat3}\right)}}
\bigg[\begin{array}{l}
\text{rational expression depending only} \\
\text{on $\left(\boldsymbol{\hat z}\boldsymbol{\hat3}\right)$ that vanishes identically}
\end{array}\bigg]
= 0.\tag*{\qed}
\end{gather*}\renewcommand{\qed}{}
\end{proof}

In \cite{routh} and \cite{gray} a quadratic integral of motion $D^2$ is taken because $D^2$ enters
the expression \eqref{eq:sepeq} for the ef\/fective potential $V(\theta;D,\lambda)$. It seems
natural, however, to speak about linear integral $D=I_3\omega_3\sqrt{d(\theta)}$, since it is well def\/ined due
to $d(\theta)>0$. Any axially symmetric rolling rigid body is integrable
\cite{chaplygin,kule,zob} and it admits, beside
energy $E$, a 2-parameter family of integrals of motion depending linearly on $\omega$.
These integrals are def\/ined
through transcendental functions satisfying a certain linear 2nd
order ODE with variable coef\/f\/icients that depend on the convex
shape of the rigid body. The special feature of the rolling sphere
integrals is that they are expressed explicitly through
elementary functions.

In \cite{gray} authors also discuss the limit $R\rightarrow 0$ of the
Jellett's and Routh integrals and they f\/ind they are equivalent to conservation
of the vertical $\boldsymbol L \boldsymbol{\hat z}$ and the axial
$\boldsymbol L \boldsymbol{\hat 3}$ components of the angular momentum as for symmetric top with a f\/ixed point tip.
A slightly dif\/ferent way of f\/inding the integral $D$ has been given
in \cite{kule}. The approach of \cite{kule} captures
also the situation when the sphere is rolling along another sphere.

\section{Separation equations for rTT}

Equations \eqref{eq:theequations} can be considered as a fourth
order system for $\theta$, $\dot\theta$, $\dot\varphi$, $\omega_3=\dot\psi
+ \dot\varphi\cos\theta$ since~$\varphi$,~$\psi$ are cyclic
variables, but it is totaly a sixth order system of equations for
the Euler angles $(\theta,\varphi,\psi)$. The existence of three
integrals of motion reduces the dif\/ferential order by three and we
obtain the system of three equations
\begin{subequations}\label{eq:constmotionrTT}
\begin{gather}
D  = \omega_3\big[I_1I_3 + mR^2I_3\left( \alpha - \cos\theta \right)^2 + mR^2I_1\sin^2\theta\big]^
\frac{1}{2} = I_3\omega_3 \sqrt{d(\theta)},\label{eq:routh} \\
\lambda  = I_1\dot\varphi\sin^2\theta-\left( \alpha - \cos\theta \right) I_3\omega_3,\label{eq:jellett} \\
E  = \frac{1}{2}mR^2\big[ \dot\theta^2\left( \alpha - \cos\theta \right)^2 +
\sin^2\theta\dot\varphi^2\left( \alpha - \cos\theta \right)^2 +
2\sin^2\theta\dot\varphi\omega_3\left( \alpha - \cos\theta \right)
 \nonumber \\
\phantom{E  =}{} +
\sin^2\theta\omega_3^2 +
\dot\theta^2\sin^2\theta\big] + \frac{1}{2}\big[ I_1\dot\varphi^2\sin^2\theta + I_1\dot\theta^2 +
I_3\omega_3^2\big] + mgR\left( 1 - \alpha\cos\theta\right).
 \label{eq:energy}
\end{gather}
\end{subequations}
It is separable when we resolve \eqref{eq:jellett} for
$\dot\varphi = \frac{\lambda + \left( \alpha - \cos\theta \right)
I_3\omega_3}{I_1\sin^2\theta} $, substitute $\dot\varphi$ into the
energy integral~\eqref{eq:energy} and f\/inally express $\omega_3$
in~\eqref{eq:energy} through $D$ and $\theta$, using the Routh
integral~\eqref{eq:routh}. Then we obtain a separable equation of
the form
\begin{equation*}
E = g(\cos\theta)\dot\theta^2 + V(\cos\theta,D,\lambda)
\end{equation*}
with
\begin{gather}
g(\cos\theta) =
\frac{1}{2}I_3\big( \beta\big(\left(\alpha-
\cos\theta\right)^2+1-\cos^2\theta\big) + \gamma\big)  \nonumber \\
 V(z,D,\lambda) = mgR\left( 1-\alpha z\right) + \frac{1}{2I_3d(z)} \left[
\frac{\big(\lambda\sqrt{d(z)} + \left(\alpha-z\right) D \big)^2
\big(\beta\left(\alpha-z\right)^2 + \gamma\big)}
{\gamma^2\left( 1-z^2\right)} \right.\nonumber  \\
\left.\phantom{V(z,D,\lambda) =}{} + D^2\left(\beta\left( 1-z^2\right) + 1\right) + \frac{2D\beta\left(
\alpha-z\right)\big(\lambda\sqrt{d(z)} + \left(\alpha-z\right) D \big)}{\gamma}
 \right]. \label{eq:sepeq}
\end{gather}
We denote $z=\cos\theta$, $\gamma = \frac{I_1}{I_3}$,
$\beta = \frac{mR^2}{I_3}$ and
$d(z)=\gamma + \beta\left(\alpha-z\right)^2 + \gamma\beta\left( 1-z^2\right)$.

Notice that both $g(z)>0$ and $d(z)>0$
are strictly positive functions
for $z\in [-1,1]$ ($\theta\in [0,\pi]$).
The potential $V(z=\cos\theta,D,\lambda)$ is well def\/ined for all values
of $\theta \in ]0,\pi[$ and
$V(\cos\theta,D,\lambda) \rightarrow +\infty$,
$\theta\rightarrow 0,\pi$ if
the quotient of integrals of motions
$\frac{D}{\lambda} \neq \frac{-1}{\left(\alpha \pm 1\right)}
\big( \gamma+\beta\left(\alpha \pm 1\right)^2\!\big)^\frac{1}{2}$.

Since $V(\theta,D,\lambda)$ has no singularities in $\theta \in ]0,\pi[$,
the motion is conf\/ined to the interval
determined by the equation $\dot\theta = 0$ which gives condition
$E = V(\cos\theta,D,\lambda)$.
For f\/ixed values of~$\lambda$,~$D$ and for the energy
$E \geq V(z_{\min} = \cos\theta_{\min},D, \lambda)$ the equation
$E = V(\cos\theta,D, \lambda)$ has, at least,
two solutions $\theta_1$ and $\theta_2$ and the one-dimensional
$\theta$-motion takes place between two turning points
$0 \leq \theta_1, \theta_2 \leq \pi$. The motion of the
symmetry axis $\boldsymbol{\hat3}$ on the unit sphere $S^2$ takes
place between two latitudes
and has nutational character similarly
as the nutational motion of the Lagrange top (LT).

For $\frac{D}{\lambda} = \frac{-1}{\left(\alpha \pm 1\right)}
\big( \gamma+\beta\left(\alpha \pm 1\right)^2\big)^\frac{1}{2}$ the potential
takes the form
\begin{gather*}
 V\left( z,D=\frac{-\lambda}{\left(\alpha\pm1\right)}
\big(\gamma + \beta\left(1\pm\alpha\right)^2\big)^\frac{1}{2}, \lambda\right) =
mgR\left(1-\alpha z\right) +\frac{\lambda^2}{2I_3d(z)\left(\alpha\pm1\right)^2}  \\
\qquad{}\times\left[\big( \gamma +
\beta\left(\gamma\pm1\right)^2\big)\left(\beta\left(1-z^2\right)+1\right) +
\frac{\big(\sqrt{d(z)}-\left(\alpha-z\right)\big(\gamma +
\beta\left(\gamma\pm1\right)^2\big)^\frac{1}{2}\big)^2}{\gamma^2\left(1-z^2\right)}\right.  \\
 \left. \qquad{}-
\frac{\big(\sqrt{d(z)}\left(\alpha\pm1\right) - \left(\alpha-z\right)\big(\gamma +
\beta\left(1\pm\alpha\right)^2\big)^\frac{1}{2}\big)}
{\gamma}2\beta\left(\alpha-z\right) \left(\gamma +
\beta\left(1\pm\alpha\right)^2\right)^\frac{1}{2} \right].
\end{gather*}

In order to discuss the motion of the rTT we shall revoke
similarities of equations \eqref{eq:constmotionrTT} with the
equations of the Lagrange top (LT) that follow from the Lagrangian \cite{ll}
\begin{equation*}
\mathscr{L} = \frac{1}{2}I_1\big(\dot\theta^2+\dot\varphi^2\sin^2\theta\big) +
\frac{1}{2}I_3\big(\dot\psi+\dot\varphi\cos\theta\big)^2 -
mgl\cos\theta,
\end{equation*}
where the angles $\theta$, $\varphi$ and $\psi$ has the same meaning
as for TT. The Lagrange equations of the LT admit the following three
integrals of motion
\begin{subequations}\label{eq:constmotionHST}
\begin{gather}
L_3  =  I_3\big(\dot\psi + \dot\varphi\cos\theta\big) = I_3\omega_3, \\
L_z  =  I_1\dot\varphi\sin^2\theta + I_3\cos\theta
\big(\dot\psi+\dot\varphi\cos\theta\big) =
I_1\dot\varphi\sin^2\theta + I_3\omega_3\cos\theta, \\
E  =  \frac{1}{2}I_1\big(\dot\theta^2+\dot\varphi^2\sin^2\theta\big) +
\frac{1}{2}I_3\big(\dot\psi+\dot\varphi\cos\theta\big)^2 +
mgl\cos\theta,
\end{gather}
\end{subequations}
where $L_3$, $L_z$ and $E$ denote the constant values of
integrals. There are transparent similarities between equations
\eqref{eq:constmotionrTT} and \eqref{eq:constmotionHST}. The Routh
integral $D$ corresponds to $L_3 = I_3\omega_3$, the Jellett
integral has a similar structure as $L_z$ and in the energy $E$
both $\dot\varphi$ and $\omega_3$ can be eliminated to give a
one-dimensional equation for $\theta(t)$. However in the case of
rTT the ef\/fective potential $V(z,D,\lambda)$ becomes a more
complicated function of $z$ than the LT ef\/fective potential
$V_{LT} = \frac{\left( L_z - L_3\cos\theta\right)^2}{I_1^2\sin^2\theta} +
\frac{2mgl}{I_1}\cos\theta $. We shall analyze the character of
motions of the rTT by studying some special types of solutions
that are natural counterparts of special solutions to the LT.

In the case of LT one can distinguish several types of special solutions
def\/ined by an invariant algebraic conditions for dynamical variables and/or
by f\/ixing values of integrals of motion. Their trajectories are customarily
represented by a curve on $S^2$ drawn by the symmetry axis $\boldsymbol{\hat3}$ of LT.
\begin{itemize}\itemsep=0pt
\item[a)]
Vertical rotations are def\/ined by the condition $\theta(t) \equiv 0$
or $\theta(t) \equiv \pi$, so that $L_z=\pm L_3=
I_3(\dot\psi \pm \dot\varphi)$.
The vertical rotations are represented by the north and
the south pole on $S^2$.
\item[b)]
Planar pendulum motions have $L_3 = I_3\omega_3=0$,
$L_z = I_1\dot\varphi\sin^2\theta + I_3\omega_3\cos\theta=0$
so that either $\omega_3 = \dot\varphi=0$ or $\omega_3 = 0$
and $\theta(t) \equiv 0,\pi$. They move along large circle arcs
through the south pole of $S^2$.
\item[c)]
Spherical pendulum motions have $L_3 = I_3\omega_3=0$,
$L_z = I_1\dot\varphi\sin^2\theta + I_3\omega_3\cos\theta \neq 0$
so that $\omega_3 = 0$ and $L_z = I_1\dot\varphi\sin^2\theta$.
The $\dot\varphi$ does not change the sign and the trajectory
$(\varphi(t),\theta(t))$ of $\boldsymbol{\hat3}$ draws a wavelike curve
between two latitudes $0<\theta_1<\theta_2<\pi$ determined
by the condition $E=\frac{1}{2}\frac{L_z^2}{I_1\sin^2\theta}
+mgl\cos\theta$.
\item[d)]
Precessional motions with $\dot\theta = 0$ which implies that
$\theta(t) \equiv \theta_0$, $0<\theta_0<\pi$ and
that $\dot\psi$ and~$\dot\varphi$ are also constant.
They are represented by latitude circles on $S^2$.
\item[e)]
General nutational motions between two latitudes
$0 \leq \theta_1 \leq \theta_2 \leq \pi$ given as solutions
of the equation $E =
\frac{1}{2I_1}\frac{\left( L_z - L_3\cos\theta\right)^2}{\sin^2\theta} +
\frac{1}{2I_3}L_3^2 + mgl\cos\theta$.
As is well known, the shape of the curve drawn by $\boldsymbol{\hat3}$ on the unit
sphere depends on whether $\dot\varphi = \frac{L_z - L_3\cos\theta}
{I_1\sin^2\theta}$ changes sign during motion or not.
There are three cases:
\begin{itemize}\itemsep=0pt
\item[(i)] $\frac{L_z}{L_3} \notin [\cos\theta_2,\cos\theta_1]$
so that  $\dot\varphi$ has the same sign all the time and the
trajectory $(\varphi(t),\theta(t))$ of $\boldsymbol{\hat3}$ on $S^2$
has a wavelike form.
\item[(ii)]  $\frac{L_z}{L_3} = \cos\theta_1$.
In this case $\dot\varphi$ will be zero at $\theta_1$ and the curve
$(\varphi(t),\theta(t))$ has cusps at the latitude $\theta_1$.
\item[(iii)] $\frac{L_z}{L_3} \in ]\cos\theta_2,\cos\theta_1[$
so that $\dot\varphi$ changes sign during motion and has dif\/ferent signs
at the both turning angles $\theta_1$ and $\theta_2$.
The axis $\boldsymbol{\hat3}$ draws a wave like curve with loops on~$S^2$.
\end{itemize}
\end{itemize}

\begin{figure}[!h]
\centerline{\includegraphics{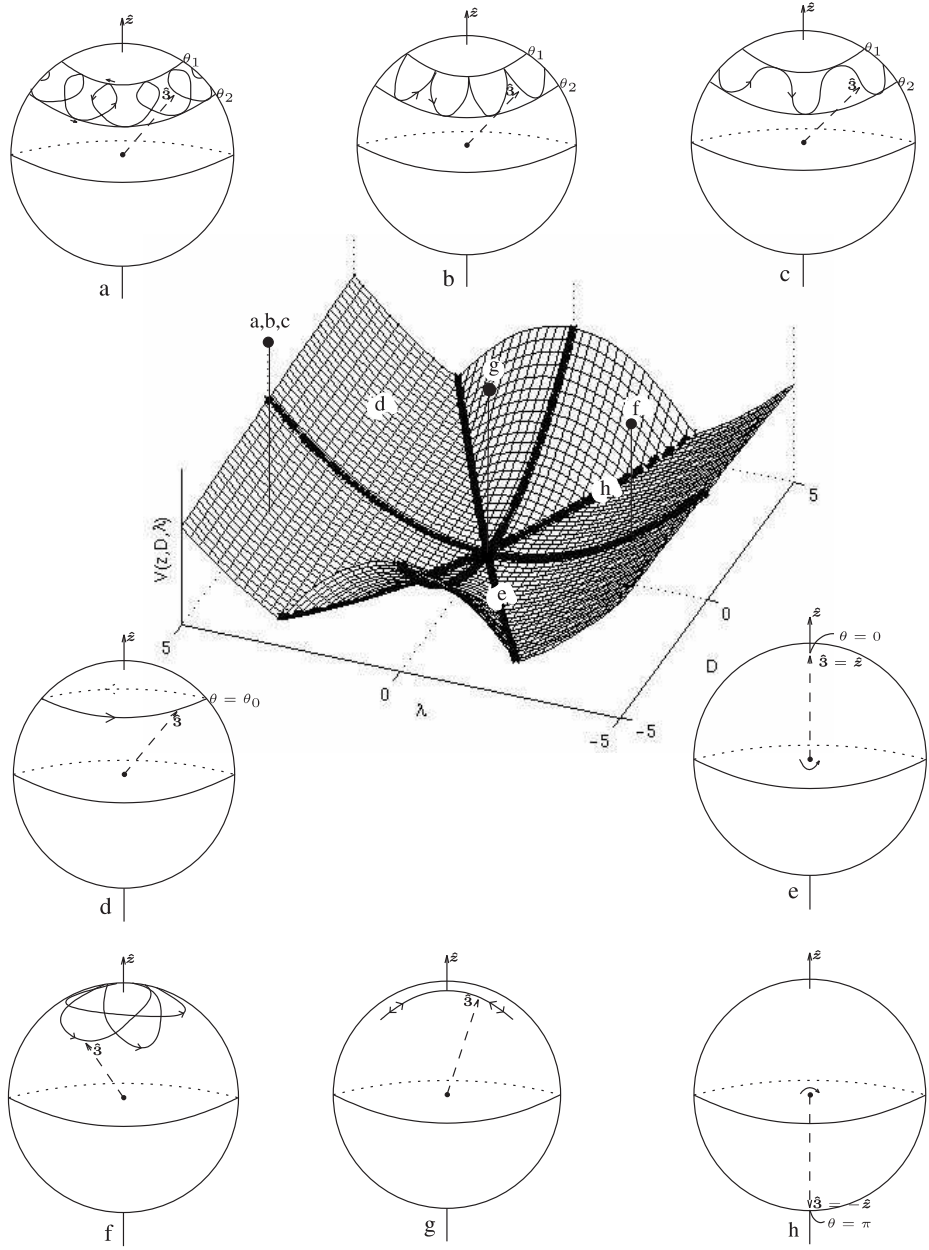}}
\caption[Illustration of the phase space picture of
TT]{Illustration of the phase space picture of TT.}
\label{fig:phasespace}
\end{figure}

Due to similarity of separation equations
\eqref{eq:constmotionrTT} with \eqref{eq:constmotionHST} we can
distinguish for rTT similar types of invariant solutions as for
LT. They are easier to analyze than the general case $E =
g(\cos\theta) \dot\theta^2 + V(\cos\theta,D,\lambda)$ and they
well illustrate behavior of rTT for dif\/ferent initial conditions.
They are again represented by curves drawn by the symmetry axis
$\boldsymbol{\hat3}$ on $S^2$ (see Fig.~\ref{fig:phasespace}).
\begin{itemize}\itemsep=0pt
\item[a)] Vertical rotations def\/ined by the condition $\theta(t)
\equiv 0$ or $\theta(t) \equiv \pi$. As for LT, the vertical
rotations are represented by the north and the south pole on
$S^2$. But here for vertical motions $\frac{D}{\lambda} =
\frac{-1}{(\alpha \mp 1)} \big( \gamma+\beta\left(\alpha \mp
1\right)^2\big)^\frac{1}{2}$ for $\theta = 0,\pi$ respectively.
\item[b)] Analog of planar pendulum type of solutions: $D =
I_3\omega_3\sqrt{d(\theta)}=0$, $\lambda =
I_1\dot\varphi\sin^2\theta - \left( \alpha - \cos\theta \right) I_3\omega_3=0$ so that either
$\omega_3 = \dot\varphi=0$ or $\omega_3 = 0$ and $\theta = 0,\pi$.
The potential $V(z=\cos\theta,D=0,\lambda=0) =
mgR(1-\alpha\cos\theta)$ is a periodic bounded function, the
coef\/f\/icient $g(\cos\theta) = \frac{1}{2}\big[I_1 + \left( \alpha - \cos\theta \right)^2 +
1-\cos^2\theta\big] > 0$ is also a positive periodic bounded
function so that rTT admits pendulum type of solutions for low
values of energy as well as the ``rotational'' type solutions (as
in the mathematical pendulum) for higher values of energy.
\item[c)] Analog of the $L_3 \neq 0$, $L_z = 0$ case:
$\lambda =I_1\dot\varphi\sin^2\theta - \left( \alpha - \cos\theta \right) I_3 \omega_3 = 0$, $D \neq 0$
which means that $\boldsymbol L$ is always in the plane orthogonal to $\boldsymbol
a$. Then
\begin{gather*}
E =g(\cos\theta)\dot\theta^2 + V(z,D,\lambda=0) =
g(\cos\theta)\dot\theta^2 +
\frac{D^2}{2I_3d(z)}\bigg[\beta\left(1-z^2\right)   \\
\phantom{E =}{} +1+\frac{\left(\alpha-z\right)^2}{\gamma^2\left(1-z^2\right)^2}\big(
\beta\left(\alpha-z\right)^2+\gamma\big) +
\frac{2\beta\left(\alpha-z\right)^2}{\gamma}\bigg] +
mgR\left( 1-\alpha z\right)
\end{gather*}
and the potential $V(z=\cos\theta,D,\lambda = 0)
\rightarrow +\infty \text{ as }\theta \rightarrow 0,\pi$. It has exactly one
minimum for $\theta \in ]0,\pi[$ because it is a convex function of $z$.
To prove this we show that
\begin{gather*}
\frac{d^2V(z,D,\lambda=0)}{dz^2} =
\frac{-2D^2\alpha}{I_3\gamma^2\left(1-z^2\right)}
\left( z^3-\frac{3}{2}\frac{1+\alpha^2}{\alpha}z^2+3z-
\frac{1+\alpha^2}{2\alpha}\right)
\end{gather*}
is positive for all $-1<z<1$. The third order polynomial $p(z) =
z^3-\frac{3}{2}\frac{1+\alpha^2}{\alpha}z^2+3z-
\frac{1+\alpha^2}{2\alpha}$ is negative for $z=0$, and it is not
changing sign because it does not have a zero in the interval
$-1<z<1$ for $0<\alpha<1$. The derivative $p'(z) = 3\big( z^2 -
\frac{1+\alpha^2}{\alpha}z + 1\big)$ has two zeros $z_1 = \alpha$, $z_2
= \frac{1}{\alpha}$ and $p(z_1=\alpha) =
-\frac{1}{2\alpha}(\alpha^2 - 1) < 0$. Since it is a local
maximum, $p(z)$ can't change sign. The function
$V(z=\cos\theta,D,0)$ is convex and has only one minimum.
\item[d)] Analog of spherical pendulum solutions: $D =
\omega_3\sqrt{d(\theta)}=0$, $\lambda = I_1\dot\varphi\sin^2\theta$
$\neq 0$ so that $\omega_3 = 0$ since $d(\theta)>0$ for all $0 \leq
\theta \leq \pi$. This means that during the rolling motion of rTT
it's not performing any rotation around $\boldsymbol{\hat3}$-axis. Then
$\dot\varphi=\frac{\lambda}{I_1\sin^2\theta}$ and
\begin{gather*}
E  =  g(\cos\theta)\dot\theta^2 + V(z=\cos\theta,D=0,\lambda)  \\
\phantom{E}{} =  g(\cos\theta)\dot\theta^2 +
\frac{\lambda^2}{2I_1\left(1-z^2\right)\theta} \big( mR^2\left(\alpha-z\right)^2 +
I_1\big) + mgR\left(1-\alpha z\right).
\end{gather*}
The potential $V(z=\cos\theta,D=0,\lambda) \rightarrow +\infty
\text{ as }\theta \rightarrow 0,\pi$ and just as in the previous case it has
exactly one minimum for $\theta \in ]0,\pi[$ because it is a convex
function of $z$.
To prove this we show that for $0<\alpha<1$ and $\frac{\gamma}{\beta}>0$,
the derivative
\begin{gather*}
\frac{d^2V(z,0,\lambda)}{dz^2}  =
-\frac{2\lambda^2mR^2\alpha}{I_1^2\left(1-z^2\right)^3}
\left[ z^3 - \frac{3\big(1+\alpha^2+\frac{\gamma}{\beta}\big)}{2\alpha}z^2+
3z-\frac{1+\alpha^2+\frac{\gamma}{\beta}}{2\alpha}\right]
\end{gather*}
is positive for all $-1<z<1$. The third order polynomial $q(z) =
z^3 - \frac{3\left(1+\alpha^2+\frac{\gamma}{\beta}\right)}{2\alpha}z^2+
3z-\frac{1+\alpha^2+\frac{\gamma}{\beta}}{2\alpha}$ has three
solutions. One can show that, for $0<\alpha<1$ and
$\frac{\gamma}{\beta}>0$, two of the solutions will be complex and
the third solution will be greater than $z=1$. Because the leading
term in $q(z)$ is positive and the only real solution is greater
than one, we conclude that $q(z)$ is negative for $-1<z<1$ and
hence $\frac{d^2V(z,0,\lambda)}{dz^2}$ is positive for $-1<z<1$
and $V(z=\cos\theta,D=0,\lambda)$ is therefore convex.
\\
There are two types of orbits here:
\begin{itemize}
\item[(i)] $E = V(\cos\theta_{\min},D,\lambda)$ Then the motion is
along the horizontal circle at $\theta(t) \equiv \theta_{\min}$.
\item[(ii)] $E > V(\cos\theta_{\min},D,\lambda)$ Here the motion is
conf\/ined between the circles $0 < \theta_1 \leq \theta_2 < \pi$
which are the solutions to $E = V(\cos\theta,\lambda,D=0)$. Notice
that $\dot\varphi=\frac{\lambda}{I_1\sin^2\theta}$ has the same
sign during the motion.
\end{itemize}
\item[e)] Precessional motions: $\dot\theta = 0$ which implies
that $\theta(t) \equiv \theta_0$, $0 < \theta_0 < \pi$ with
$\omega_3$ and $\dot\varphi$ equal to their initial values. The
motion on $S^2$ is represented by the latitude circles
$(\theta_0,\varphi(t))$.
\item[f)] General nutational motion: The
trajectory on $S^2$ is conf\/ined between the two latitude circles
$(\theta_1,\varphi(t))$ and $(\theta_2,\varphi(t))$. Where $0 \leq
\theta_1 \leq \theta_2 \leq \pi$ are given as solutions of the
equation $E = g(\cos\theta)\dot\theta^2 +
V(z=\cos\theta,D,\lambda)$. The shape of the curve drawn by
$\boldsymbol{\hat3}$ on the unit sphere depends on whether $\dot\varphi =
\frac{D}{I_1\sin^2\theta}\big[
\frac{\lambda}{D}+\frac{\alpha-\cos\theta}{\sqrt{d(\theta)}}\big]$
changes sign during the motion or not. There are three
qualitatively dif\/ferent cases because the function $h(z) =
\frac{\left(\alpha - z\right)}{ \sqrt{d(z)}}$ is monotone when, $-1 < z <
1$. To see that $h(z)$ is monotone observe that $\frac{dh(z)}{dz}
= \frac{\gamma\left(1+\beta-\beta\alpha z\right)} {(d(z))^\frac{3}{2}}=0$
implies $z=\frac{1}{\alpha} + \frac{1}{\alpha\beta}>1$.
\begin{itemize}\itemsep=0pt
\item[(i)] $\frac{\lambda}{D} + \frac{\left( \alpha - \cos\theta \right)}{
\sqrt{d(\theta)}}\neq 0$ for $\theta \in [\theta_1,\theta_2]$, so
that $\dot\varphi$ has the same sign during the motion, and the
trajectory $(\theta(t),\varphi(t))$ has a wavelike form touching
tangentially the two boundary latitude circles $\theta_1$ and
$\theta_2$. \item[(ii)] $\frac{\lambda}{D} + \frac{\left( \alpha - \cos\theta \right)}{
\sqrt{d(\theta)}}=0$ for $\theta = \theta_1$. In this case
$\dot\varphi$ is zero at $\theta_1$, and both $\dot\theta$ and
$\dot\varphi$ vanishes at $\theta(t) = \theta_1$ the motion
momentarily stops, and the trajectory $(\theta(t),\varphi(t))$
will have cusps at the latitude $\theta_1$. \item[(iii)]
$\frac{\lambda}{D} + \frac{\left( \alpha - \cos\theta \right)}{ \sqrt{d(\theta)}}=0$ for
certain $\theta \in ]\theta_1,\theta_2[$, so that $\dot\varphi$
changes sign during the motion, and is positive at the latitude
$\theta_2$ and negative at the latitude $\theta_1$. The trajectory
$(\theta(t),\varphi(t))$ of the $\boldsymbol{\hat3}$-axis on the unit
sphere $S^2$ will have a wavelike form with loops when touching
tangentially the latitude circle $\theta_1$.
\end{itemize}
\end{itemize}
All these particular types of motion are illustrated in
Fig.~\ref{fig:phasespace} presenting the minimal surface of the
function $V(z=\cos\theta)$ in the space of parameters
$(D,\lambda,E)$. This surface is the lower boundary of the set of
all admissible values of $(D,\lambda,E)$. The values of parameters
on the minimal surface corresponding to the special types of
solutions are marked by distinguished black lines and the
corresponding motions on the sphere $S^2$ are illustrated by the
adjacent pictures marked with the same letter.
All triples of parameters $(D,\lambda,E)$  describing points on the
minimal surface def\/ine precessional motions $\theta=\theta_0={\rm const}$.
They are denoted by the letter d. The degenerate forms of precessional
motions are the vertical rotations $\theta_0=0$ and $\theta_0=\pi$
that correspond to points on the lines e and h.
All generic points above the minimal surface of $V(z)$ describe
nutational motions denoted here by letters a, b, c. Two types of
special motions corresponding to points above the minimal surface
have been distinguished: the spherical pendulum type solutions that
are marked by the letter f above the line $D=0$ and the planar
pendulum type solutions that are marked by the letter g above the
point $D=0$, $\lambda=0$.
The admissible set in the space of parameters $(D,\lambda,E)$
represents all possible
dynamical states of rTT modulo the choice of the initial angles
$\varphi_0$, $\psi_0$ of the cyclic variables and of the initial
value of the angle $\theta_0$. The time dependence of $\theta(t)$,
$\varphi(t)$, $\psi(t)$ is, up to these translations, fully determined
by each admissible point in Fig.~\ref{fig:phasespace}.

\section{Final remarks}

Separability of the rTT has been recognized many times in the
literature \cite{routh,chaplygin,gray} since the Routh
observation that the system \eqref{eq:TTnewtondyn} admits three
independent integrals of motion and is separable. There are
several studies of stationary motions of the rTT
\cite{routh,chaplygin,chap2,karapetyan}. The separation equations,
however, has been less used for analysis of the rTT and their apparent
similarity with the equations of the Lagrange top, seemingly has
not been discussed before.

In this paper we have discussed thoroughly the vector equations of
the sliding tippe top and their reduction for pure rolling motion
in the plane, the rTT equations
(\ref{eq:eq2prop1}), (\ref{eq:eq3prop1}). The equations of motion
(\ref{eq:eq2prop1}), (\ref{eq:eq3prop1})
admit three integrals of motion that are given here
both in the vector form and also in the coordinate form, which provides
the separation equations. Two integrals $(D,\lambda)$, of angular
momentum type, depend linearly on the momenta. Their existence can
be understood as ref\/lection of the fact that the dynamical
equations \eqref{eq:theequations} have two cyclic variables~$\varphi$,~$\psi$~\cite{zob}
although we are not speaking here about the
Hamilton--Jacobi type separability. The third integral is the
energy $E$ that depends quadratically on momenta and gives rise to
the main separation equation similarly as in the case of the
Lagrange top.

According to the old result by Chaplygin \cite{chaplygin} all
axially symmetric rigid bodies rolling on the plane admit three
integrals of motion -- the energy and two integrals linear in
momenta and are in principle separable. In general the linear integrals are however
 def\/ined through two particular solutions of a second order
linear dif\/ferential equation with nonelementary solutions. For
instance in the case of the rolling disc they are given by
Legendre functions. These equations allow to formally express
$\dot\varphi(\cos\theta)$ and $\omega_3(\cos\theta)$ as known
functions of $\cos\theta$ to be substituted into the energy
integral to give the $\theta$-separation equation similarly as in
this paper.

The special feature of the rTT equations is that the expressions
for $\dot\varphi(\cos\theta)$ and $\omega_3(\cos\theta)$ are
elementary functions
of $\cos\theta$. When substituted explicitly into the energy
integral they give an elementary expression for the ef\/fective
potential $V(\cos\theta)$. This simplif\/ies
analysis of equations and enables representing all possible motions
in the space of parameters $(D,\lambda,E)$. Only two types of
stationary motion describe stable trajectories with the reaction
force orthogonal to the plane. They are the stable asymptotic
motions of the genuine tippe top, of vertical spinning type and
the tumbling solutions when the center of mass is f\/ixed in the
space and the sphere is rolling along a circle.

The approach presented here makes it possible to perform more general
study of separability and of elementary separability (when
$\dot\varphi(\cos\theta)$ and $\omega_3(\cos\theta)$ can be
expressed by elementary functions) of axially symmetric rigid
bodies. These questions are currently under study and the results
will be presented in a subsequent paper.

\subsection*{Acknowledgments}
The authors would like to thank referees for useful suggestions and pointing some references.

\pdfbookmark[1]{References}{ref}
\LastPageEnding


\begin{thebibliography}{99}

\footnotesize\itemsep=0pt

\bibitem{bourabee}
 Bou-Rabee N.M., Marsden J.E., Romero L.A.,
Tippe top inversion as a dissipation-induced instability,
{\it SIAM J. Appl. Dyn. Syst.} {\bf 3} (2004), 352--377.

\bibitem{chaplygin}
 Chaplygin S.A.,  On motion of heavy rigid body of
revolution on horizontal plane, {\it Proc. of the Physical Sciences,
Section of the Society of Amateurs of Natural Sciences} {\bf 9} (1897), no.~1,
10--16.

\bibitem{chap2}
 Chaplygin S.A.,
On a ball's rolling on a horizontal plane,
{\it Regul. Chaotic Dyn.} {\bf 7} (2002), 131--148.

\bibitem{cohen}
 Cohen R.J.,
The tippe top revisited,
{\it Amer. J. Phys.} {\bf 45} (1977), 12--17.

\bibitem{ebenfeld}
 Ebenfeld S., Scheck F.,
A new analysis of the tippe top: asymptotic states and Liapunov stability,
{\it Ann. Physics} {\bf 243} (1995), 195--217, \href{http://arxiv.org/abs/chao-dyn/9501008}{chao-dyn/9501008}.

\bibitem{gray}
 Gray C.G., Nickel B.G,
Constants of motion for nonslipping tippe tops and other tops with rounded pegs,
{\it Amer. J. Phys.} {\bf 68} (1999), 821--828.

\bibitem{karapet2}
 Karapetyan A.V.,
Qualitative investigation of the dynamics of a top on a plane with friction,
{\it J. Appl. Math. Mech.} {\bf 55} (1991), 563--565.

\bibitem{karapetyan}
 Karapetyan A.V.,
On the specif\/ic character of the application of Routh's theory to systems with dif\/ferential constraints,
{\it J. Appl. Math. Mech.} {\bf 58} (1994), 387--392.

\bibitem{kule}
 Kuleshov A.S.,
On the generalized Chaplygin integral,
{\it Regul. Chaotic Dyn.} {\bf 6} (2001), 227--232.

\bibitem{ll}
 Landau L.D., Lifshitz E.M.,
Mechanics, Pergamon Press Ltd, Oxford, 1976.

\bibitem{mosh}
 Moshchuk N.K.,
Qualitative analysis of the motion of a rigid body of revolution on an absolutely rough plane,
{\it J. Appl. Math. Mech.} {\bf 52} (1988), 159--165.

\bibitem{rauchglad}
 Rauch-Wojciechowski S., Sk\"oldstam M., Glad T.,
Mathematical analysis of the tippe top,
{\it Regul. Chaotic Dyn.} {\bf 10} (2005), 333--362.

\bibitem{routh}
 Routh E.J., The advanced part of a treatise on the
dynamics of a system of rigid bodies, Dover Publications, New
York, 1905, 131--165.

\bibitem{zob}
 Zobova A.A., Karapetyan A.V.,
Construction of Poincar\'e--Chetaev and Smale bifurcation
diagrams for conservative nonholonomic systems with symmetry,
{\it J. Appl. Math. Mech.} {\bf 69} (2005), 183--194.

\end{thebibliography}
\end{document}